\documentclass[11pt,graphicx,amsmath]{article}
\usepackage{amsmath}
\usepackage{graphicx}
\usepackage{bm}
\usepackage[dvips]{color}

\usepackage{color}

\def\gsim{\;\rlap{\lower 2.5pt  \hbox{$\sim$}}\raise 1.5pt\hbox{$>$}\;}
\def\lsim{\;\rlap{\lower 2.5pt  \hbox{$\sim$}}\raise 1.5pt\hbox{$<$}\;}
\def\edth{\;\raise1.0pt\hbox{$'$}\hskip-6pt\partial\;}
\def\baredth{\;\overline{\raise1.0pt\hbox{$'$}\hskip-6pt \partial}\;}

\def\be{\begin{equation}}
\def\ee{\end{equation}}
\def\ba{\begin{eqnarray}}
\def\ea{\end{eqnarray}}

\title{ A class of transient acceleration models consistent with Big Bang Cosmology}

\author{\small Tianlong Zu \thanks{zutl@mail.ustc.edu.cn },
               Jiewen Chen \thanks{chjw@mail.ustc.edu.cn },
               Yang Zhang  \thanks{yzh@ustc.edu.cn}\ \\
     \small \emph{Key Laboratory for Researches in Galaxies and Cosmology}, \\
     \small \emph{Department of  Astronomy,  University of Science and Technology of China}, \\
     \small \emph{Hefei, Anhui, 230026,  China}\\     }

\date{}

\begin{document}
\maketitle

\begin{abstract}
{
Is it possible that the current cosmic accelerating expansion
will turn into a decelerating one?
Can this transition be realized by some viable theoretical model
that is consistent with the standard Big Bang cosmology?
We study a class of phenomenological models of a transient acceleration,
based on a dynamical dark energy
with a very general form of equation of state $p_{de}=\alpha\rho_{de}-\beta\rho_{de}^m$.
It mimics the cosmological constant $\rho_{de}\rightarrow$
const for small scale factor $a$,
and behaves as a barotropic gas with
$\rho_{de}\rightarrow a^{-3(\alpha+1)}$ with $\alpha\ge 0$
for large $a$.
The cosmic evolution of four models in the class has been examined in details,
and all yields a smooth transient acceleration.
Depending on the specific model,
the future universe may be dominated either by the dark energy or by the matter.
In two models,
the dynamical dark energy can be explicitly realized by a scalar field
with an analytical potential $V(\phi)$.
Moreover,
the statistical analysis shows that
the models can  be as robust as $\Lambda$CDM
in confronting the observational data of SN Ia, CMB, and BAO.
As improvements over the previous studies,
our models overcome the over-abundance problem of dark energy
during early eras,
and satisfy the constraints on the dark energy from WMAP observations of CMB.
}
\end{abstract}


\begin{center}
\textbf{1.Introduction}
\end{center}

Cosmological observations,
such as the SN Ia \cite{Riess,Permutter}
and CMB anisotropies \cite{WMAP Collaboration} ,
have indicated  that the  universe  is now   in an  accelerating expansion.
Interpreting in the framework of general relativity,
the acceleration can be attributed to the existence
of some dark energy, which currently dominates the total cosmic energy in the Universe.
There have been a number of candidates for the dark energy,
such as the cosmological constant $\Lambda$,
various scalar field dynamical dark energy models,
 \cite{Ratra,Wands,Barreiro,Liddle,M.L. Tong,Copeland},
Chaplygin gas model \cite{Kamenshchik},
Quantum Yang-Mills condensate models \cite{Zhang,SWang,Fu}, etc.
So far,
there is no observational evidences whether the current acceleration
is eternal or transient.
In an eternally accelerating universe,
there is an event horizon, and the S-matrix in the string theory
will be ill-defined  \cite{Hellerman, Fischler, Witten}.
Recently, the analysis of combined data of SN Ia+BAO+CMB  \cite{Shafieloo}
seems to indicate that the acceleration may be slowing down.
More data are needed for the issue \cite{Guimares}.
Thus the possibility of return to decelerating expansion in the future
has been explored.
There have been various models based on
different possible mechanisms
\cite{Townsend,SahniShtanov,Russo,Srivastava,Qiang Wu,Yungui Gong,Bento,Pavon}.
In particular,
a scalar field with an exponential potential as dark energy in Ref.\cite{Russo},
and two coupled scalar fields as dark energy in \cite{Blais} are studied respectively,
and, for certain range of parameters,
a transient acceleration occurs in these scalar models.
Refs.\cite{Fabris,Xi-ming Chen,Costa}
consider possible interaction between matter and dark energy
that can lead to a transient accelerating expansion.
A coupling between Chaplygin gas and scalar field is studied in \cite{Bilic}.
Based certain ansatzs on the dark energy density
to achieve a return to deceleration,
some scalar field models are proposed,
which have a potential of exponential type
with a quadratic dependence on the scalar field \cite{Carvalho,Alcaniz2,Alcaniz}.
However, as has been checked  \cite{Cui},
when extended back to the earlier stages,
these exponential scalar
dark energy would be dominant over the matter component,
jeopardizing  the standard Big Bang cosmological scenario.

To handle this over-abundance problem within the whole class of scalar field models,
one might directly design some special form of scalar potential
and go ahead by trial and error to see if it works.
This is essentially the method in Refs.\cite{Carvalho,Alcaniz2,Alcaniz}
which has not worked for their chosen form potential.
Moreover, there are infinite number of possible forms of scalar potentials,
and it is not practical to try each of them.
Viewing this,
we adopt a parameterization approach instead.
That is, we take some simple form of dynamic dark energy density $\rho_{de}(t)$
which is subdominant to the matter density $\rho_m(t)$ during the early stage
of the cosmic expansion.
If this works, it will automatically
overcome the over-abundance problem
and provide a possible model of transient acceleration.
In this paper,
we will specifically work with those $\rho_{de}$,
which mimics the cosmological constant for a small scale factor $a$,
and behaves, for large $a$,  like  a barotropic gas with
$\rho_{de}\rightarrow a^{-3(\alpha+1)}$ with $\alpha\ge 0$.
In certain cases,
an explicit expression of scalar potential $V(\phi)$
is obtained analytically.
Depending on specific models in the class,
the future universe may be dominated either by dark energy
or by the matter.
The interesting characteristic of our simple scalar models is that
the dark energy will always be subdominant
to the matter when extended to earlier stages,
allowing for a transient acceleration within the framework
of the standard Big Bang cosmology.

\begin{center}
\textbf{2. Models}
\end{center}

The spacetime background is the homogeneous and isotropic
flat FRW metric:
\be \label{metric}
ds^2=-dt^2+a^2(t)(dx^2+dy^2+dz^2) .
\ee
We will set $a=1$ as the current value.
The dynamical expansion of spacetime is determined by
the Friedmann equation:
\be \label{friedmann eq1}
(\frac{\dot{a}}{a})^2=\frac{8\pi G}{3}(\rho_{de}+\rho_{m}),
\ee
\be \label{friedmann eq2}
\frac{\ddot{a}}{a}=-\frac{4\pi G}{3}(\rho_{m}+\rho_{de}+3p_{de}),
\ee
where $\rho_{de}$ is the dark energy density to be discussed in the following,
$\rho_{m}  =\Omega_m\rho_ca^{-3}$ is the matter density,
and   $\Omega_m+\Omega_{de}=1$.

\begin{figure}
\begin{minipage}[t]{0.5\linewidth}
\centering
\includegraphics[width=3.2in]{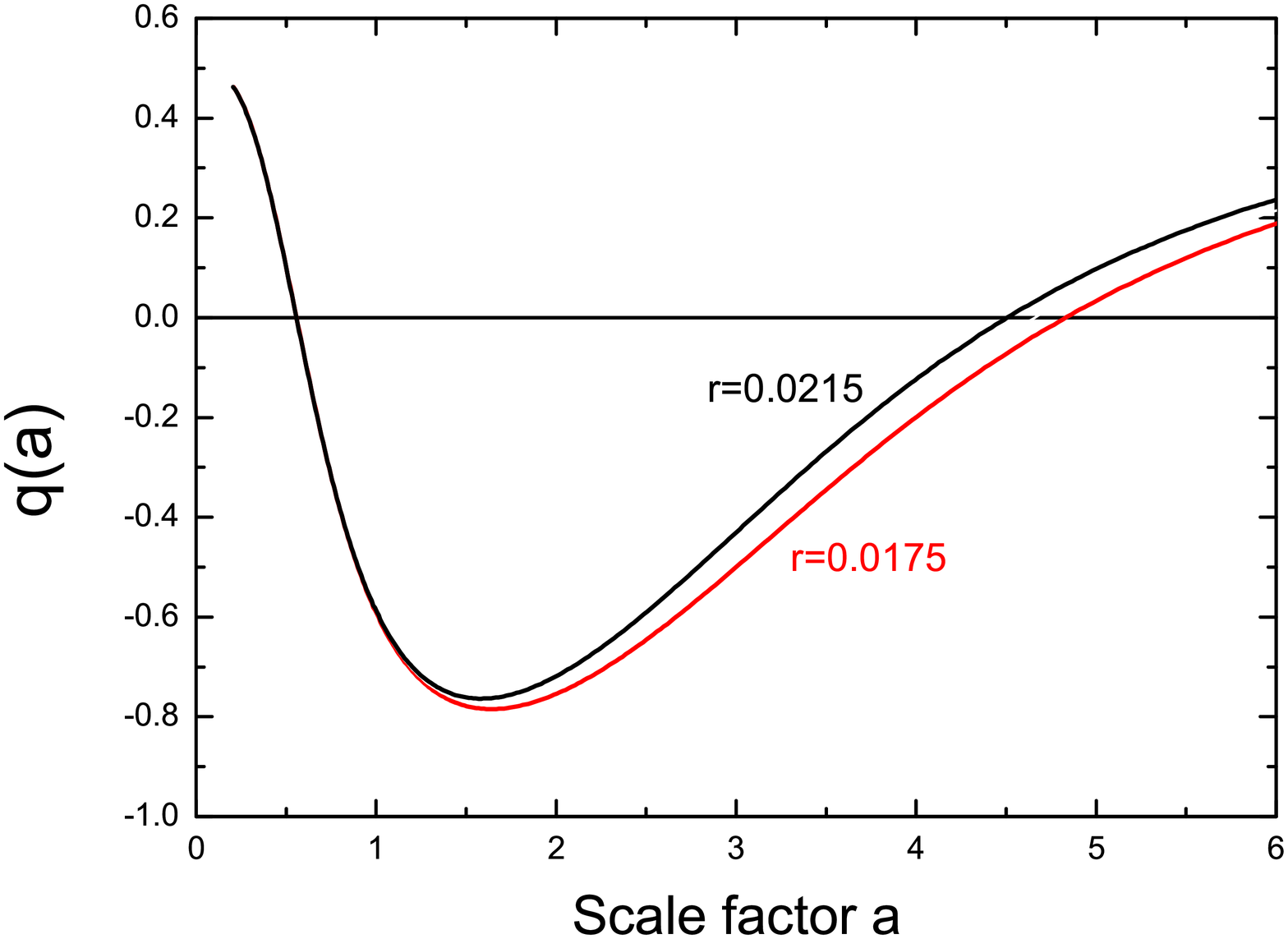}
\caption{\label{q1}
\small The deceleration parameter  $q$
in Eq.(\ref{deceleration parameter}).}
\end{minipage}
\hspace{0.5cm}
\begin{minipage}[t]{0.5\linewidth}
\centering
\includegraphics[width=3.2in]{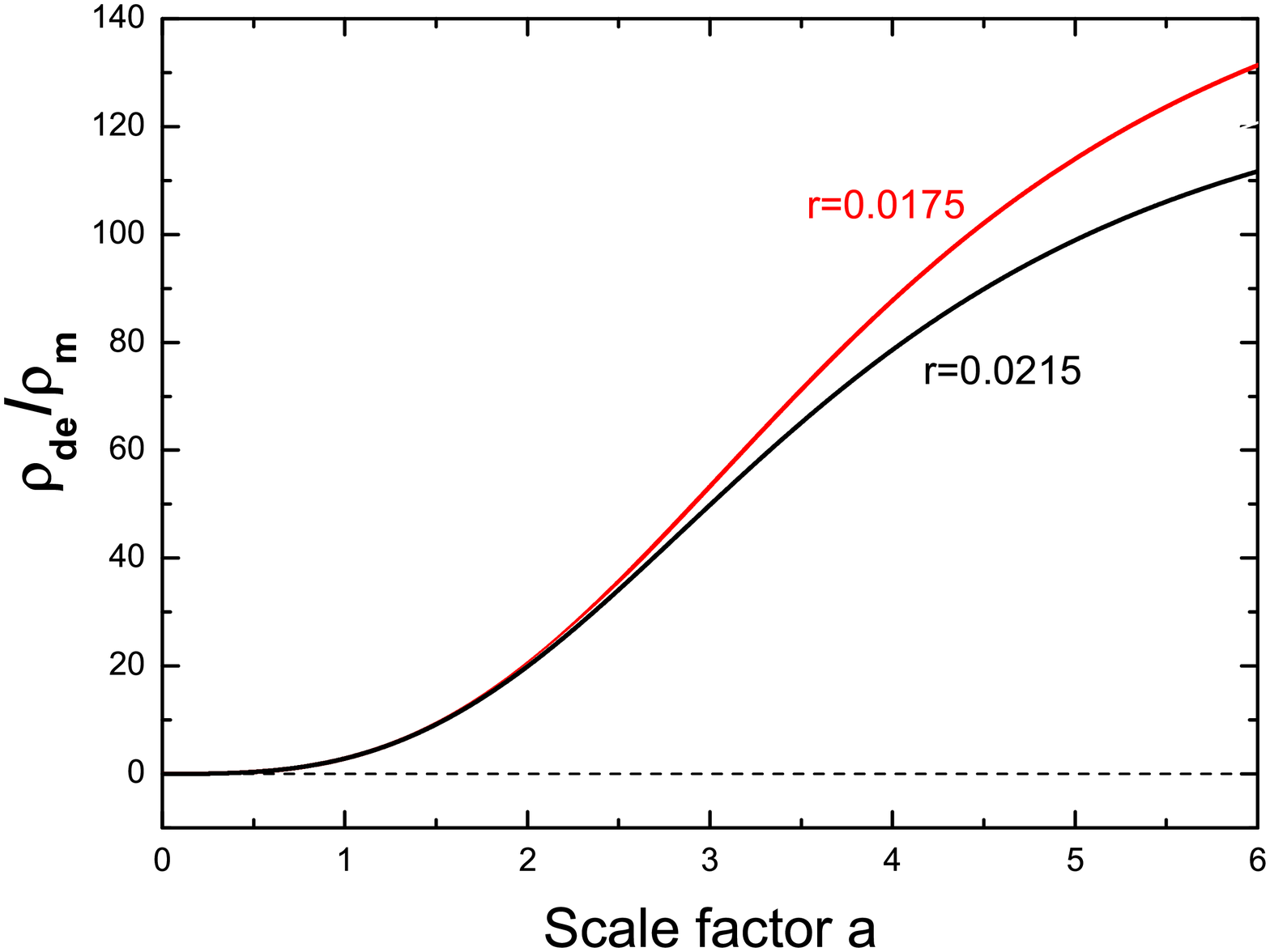}
\caption{\label{r1}
\small The ratio $\rho_{de}/\rho_{m}$
         in the model of Eq.(\ref{pressure}).
         $\rho_{de}$ will still dominate over $\rho_{m}$ in future. }
\end{minipage}
\end{figure}

First, we consider a dynamical dark energy
density as a function of the scale factor $a$:
\be \label{de density}
\rho_{de}(t)= \Omega_{de}\rho_c\frac{1+r}{1+ra(t)^3},
\ee
where $\Omega_{de}$ is the current  dark energy fraction,
$\rho_c$ is the critical density,
and $r$ is a dimensionless  parameter of the model.
It behaves as
$\rho_{de} \propto \Omega_{de}\rho_c (1+r)$ for $a^3 \ll 1/r$,
and  $\rho_{de} \propto a^{-3}$ for $a^3 \gg 1/r$,
similar to that of the matter component.
But we do not regard  $\rho_{de}$ as the matter.
For simplicity, we do not include the coupling between
the dark energy  and the matter.
By the equation of energy conservation
$a\frac{d \rho_{de}}{d a}+3(\rho_{de}+p_{de})=0$,
the pressure of the dark energy is given by
\be \label{pressure}
p_{de}(t)=-\Omega_{de}\rho_c\frac{(1+r)}{(1+ra(t)^3)^2}
         =  -\frac{\rho^{2}_{de}(t)}{\Omega_{de}\rho_c(1+r)}.
\ee
The equation of state of dark energy is
$w_{de}=p_{de}/\rho_{de}= -\frac{\rho_{de}}{\Omega_{de}\rho_c(1+r)}$.
Eq.(\ref{pressure})
turns out to be similar to a model $p \propto -\rho^{-\alpha}$, $\alpha<-1$ in Ref.\cite{Sen},
but is different from the generalized Chaplygin gas model
$p \propto -\rho^{-\alpha}$, $\alpha\geq -1$ \cite{Kamenshchik}.
From Eq.(\ref{friedmann eq1}) and (\ref{friedmann eq2}) follows the
deceleration parameter explicitly
\be \label{deceleration parameter}
q=-\frac{\ddot{a}a}{(\dot{a})^2}
=\frac{1}{2}-\frac{3\Omega_{de}(1+r)a^3}{2(1+ra^3)
[\Omega_m+(r+\Omega_{de})a^3]}.
\ee
We plot $q$ as a function of $a$ in Fig. \ref{q1}.
Here $\Omega_{de}=0.74$ and $\Omega_{m}=0.26$ are taken for concreteness.
It is seen that  $q\rightarrow \frac{1}{2}$ as its asymptotic values
in both the limits $a\rightarrow 0$ and $a\rightarrow \infty$.
Therefore,
this model predicts a decelerating expansion
for both the past and for the future,
and,  in the interval  $a\sim (0.5, 5)$, $q<0$,
the current acceleration is transient,
as shown in Fig 1 for various values of $r$.
In fact,  for a finite value of $r>0$ and a constraint
$\frac{3r}{4r+1}<\Omega_{de}$,
the acceleration is transient
and is always followed by a deceleration expansion.
By checking through calculation,
a larger $\Omega_{de}$ yields an
earlier entry into the current  acceleration,
and a larger value of parameter $r$ yields an earlier return to deceleration.
In the limiting case $r=0$,
the model reduces to  the $\Lambda$CDM model.
On the other hand, for  $r\ne 0$,
the model predicts
a dark energy fraction $\sim \Omega_{de}(1+r)$
for the early stage.
However, based on the observations of CMB,
 WMAP7 has given an error band $\pm 4\%$ for $ \Omega_{\Lambda}$
 for the $\Lambda$CDM model \cite{WMAP Collaboration}.
To be consistent with the observations,
we impose an upper limit $r \simeq 0.04$ for our model.
That is,  for the parameter $r\sim (0, 0.04)$,
our model is within the constraint from the the observations by WMAP.
In Fig. \ref{r1} we plot the ratio $\rho_{de}/\rho_m$ as a function of $a$.
At early times when $a\ll 1$,
$\rho_{de}/\rho_m \rightarrow 0$,
the dark energy is comparatively small
and the universe is matter dominated, as desired.
In the remote future when $a\gg 1$,
 the ratio $\rho_{de}/\rho_m \rightarrow (1+r)\Omega_\Lambda/r\Omega_m>1$,
asymptotically approaching a constant.
Thus the universe in the future is dominated  by $\rho_{de}$,
which is decreasing as $\propto a(t)^{-3}$,
and $a(t)\propto t^{2/3}$,
expanding like the matter-dominated.

\begin{figure}[htbp]
\centering
\includegraphics[width=0.60\columnwidth]{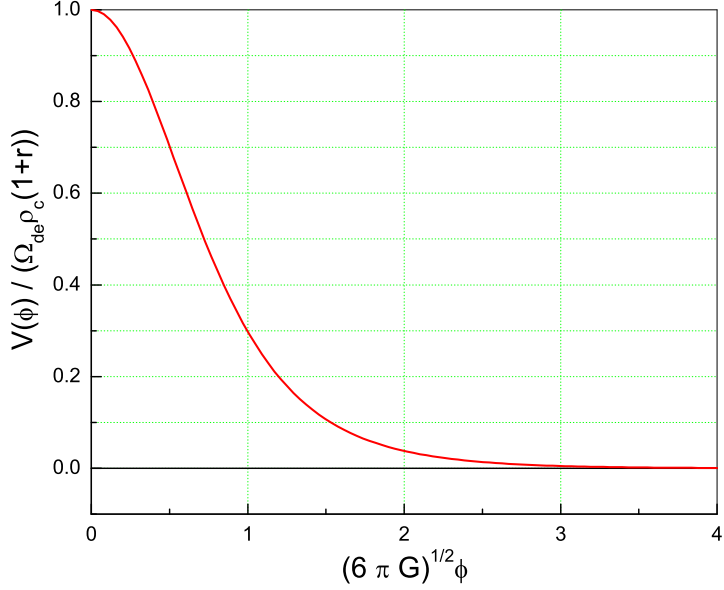}
 \caption{\label{V1}  \small
 The re-scaled potential $V(\phi)/\Omega_{de}\rho_c(1+r)$ is shown.
  }
\end{figure}

The dark energy in this model
can also be explicitly realized by a scalar field $\phi$.
For simplicity,
we do not consider the effects of matter component  \cite{Barrow}.
We  start with the
Lagrangian of the scalar field:
\be \label{lagrangian}
\mathcal{L}=\frac{1}{2}\dot\phi^2-V(\phi),
\ee
where $V(\phi)$ is the potential to be determined.
The energy density and the pressure are
$\rho_{\phi}=\frac{1}{2}\dot{\phi}^2+V(\phi)$ and
$p_{\phi}=\frac{1}{2}\dot{\phi}^2-V(\phi)$,
and satisfy the conservation equation
$a\frac{d\rho_{\phi}}{da}+3(\rho_{\phi}+p_{\phi})=0$,
which can be written as  the scalar field evolution equation:
\be \label{scalar evolution}
\dot{\phi}^2=-\frac{1}{3}a\frac{d\rho_{\phi}}{da}.
\ee
Now require that $\rho_{\phi}$ behaves as $\rho_{de}$ in Eq(\ref{de density}):
\be
\rho_{\phi}=\Omega_{de}\rho_c\frac{1+r}{1+ra^3}.
\ee
Using Eq(\ref{friedmann eq1}) without the matter,
Eq.(\ref{scalar evolution}) can be written as
\be \label{scalar evolution2}
\sqrt{\frac{8\pi G}{3}}\frac{d\phi}{da}=\pm \sqrt{\frac{ra}{1+ra^3}}.
\ee
By integrating, one obtains
\be \label{aphirelation}
\sqrt{6\pi G}(\phi-\phi_0)=\pm \ln{\frac{\sqrt{ra^3}
+\sqrt{1+ra^3}}{\sqrt{ra_0^3}+\sqrt{1+ra_0^3}}}.
\ee
where $a_0$ and $\phi_0$ are constants.
For simplicity, we set $\phi_0=0, a_0=0$.
Then Eq(\ref{aphirelation})
shrinks to:
\be \label{aphirelation2}
\sqrt{6\pi G}\phi=\pm \ln{\left[ \sqrt{ra^3}
+\sqrt{1+ra^3}  \right]}.
\ee
Taking the $+$ sign yileds
\be \label{phia1}
2\sqrt{ra^3}=e^{\sqrt{6\pi G}\phi}-e^{-\sqrt{6\pi G}\phi}.
\ee
By $V=(\rho_\phi-p_\phi)/2 $,  one has
\be \label{potential1}
V  =\frac{1}{2}\Omega_{de}\rho_c(1+r)\left[\frac{2+ra^3}{(1+ra^3)^2}\right].
\ee
Substituting  Eq(\ref{phia1}) into Eq(\ref{potential1}),
we finally obtain $V$ in terms of the  field $\phi$:
\be \label{potential2}
V(\phi)=2\Omega_{de}\rho_c(1+r) \left[\frac{1}{(e^{\sqrt{6\pi G}\phi}
+e^{-\sqrt{6\pi G}\phi})^2}
+\frac{4}{(e^{\sqrt{6\pi G}\phi}+e^{-\sqrt{6\pi G}\phi})^4} \right].
\ee
Note that $V(\phi)$ in Eq.(\ref{potential2}) is proportional to
the factor $\Omega_{de}\rho_c(1+r)$.
In Fig. \ref{V1} the re-scaled potential $V(\phi)/\Omega_{de}\rho_c(1+r)$
is plotted.
It is worthy noticing that the same
$V(\phi)$ holds if $``-"$ sign is taken in Eq.(\ref{aphirelation2}).
Indeed,  the resulting Lagrangian $\mathcal{L} $
is symmetric under $\phi \rightarrow -\phi$.
When  $\sqrt{6\pi G}\phi\gg 1$ the potential reduces to
$V(\phi) \propto e^{-2\sqrt{6\pi G}\phi}$,
a simple exponential function of $\phi$,
which is similar to what  Ratra and Peebles \cite{Ratra}
used for dark energy and dark matter in a different context.
Notice that,
 the expression of Eq.(\ref{potential2}) is a combination of
the exponential functions of  $\phi$,
but differs from the exponential potential with a quadratic $\phi^2$
in Refs. \cite{Carvalho,Alcaniz2}.
In particular, the profile of our $V(\phi)$ is more slope around $\phi=0$
than that in Refs. \cite{Carvalho,Alcaniz2}.

\begin{figure}[htbp]
\centering
\includegraphics[width=0.60\columnwidth]{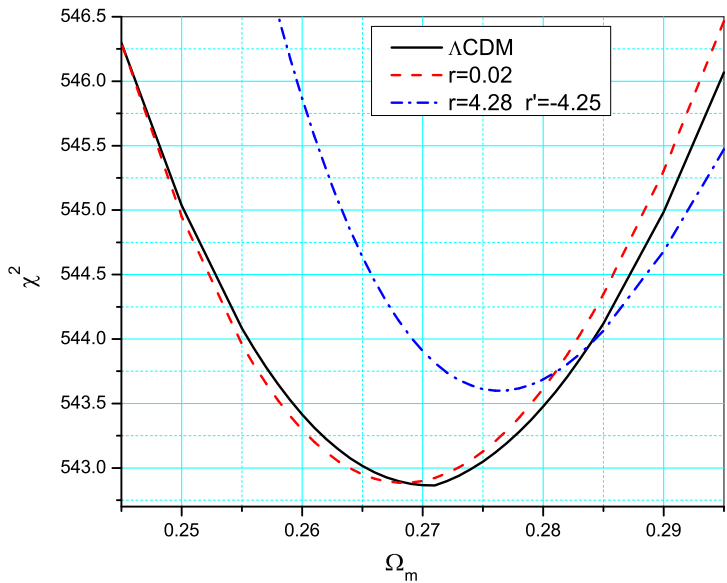}
 \caption{\label{chi}  \small
  The $\chi^2$  of the transient acceleration models
 based upon the joint data of SNIa, BAO and CMB.
 The dashed line is of the first mode with $r=0.02$
 in Eq.(\ref{de density}).
 The dotted line is of the second model
 in Eq.(\ref{density 2})  with  $r=4.28$ and $r'=-4.25$.
 For comparison, $\chi^2$  of $\Lambda$CDM ($r=0$) is also shown in the solid line.
    }
\end{figure}

To examine the observational viability of this simple model of transient acceleration,
we perform a joint analysis involving the data of SN Ia, CMB, and BAO.
We use the distance modulus $\mu_{obs}(z_i)$ data of 557 SNIa \cite{Amanullah},
the shift parameter of CMB by the WMAP observations \cite{wmap7},
and the BAO measurement from the Sloan Digital Sky Survey (SDSS) \cite{SDSS percival}.
We shall follow the calculational routine in Refs.\cite{SWang,Fu,Tong}.
Assuming that these three sets data of observations
are mutual independent and that the measurement errors for each set are Gaussian,
the likelihood function is then of the form
\begin{equation}\label{likelihood}
{\cal{L } } \propto e^{-\chi^2/2}
\end{equation}
with
\be
\chi^2 = \chi^2_{SN}+\chi_{BAO}^2+\chi_{CMB}^2.
\ee
\be\label{chi2SN}
\chi^2_{SN} =\sum_{i=1}^{557}\frac{[\mu_{obs}(z_i)-\mu_{th}(z_i)]^2}{\sigma_i^2},
\ee
\be
\chi_{BAO}^2 = \frac{(A-A_{obs})^2}{\sigma_A^2},
\ee
\be
\chi_{CMB}^2 = \frac{(R-R_{obs})^2}{\sigma_R^2}.
\ee
The detailed  specifications of these formulae have been given in Ref.\cite{Fu}.
Variations of values of the model parameters
yield respective values of $\chi^{2}$.
For demonstration, we take the model of $r=0.02$ and let $\Omega_m$ vary.
The resulting $\chi^2(\Omega_m)$ is plotted (in dash) in Fig.\ref{chi}.
To compare with the standard  $\Lambda$CDM,
we also plot the case of $r=0$ (corresponding to $\Lambda$CDM).
The minimal is $\chi^2=  542.91$ at $\Omega_m=0.268$
for the model of $r=0.02$,
whereas $\Lambda$CDM has $\chi^2=542.88$ at $\Omega_m=0.271$.
Its corresponding likelihood ${\cal{L}}$ is $0.99$ times that of $\Lambda$CDM.
Thus the joint analysis tells that the model of $r=0.02$
is quite close to $\Lambda$CDM
in confronting the observational data,
and is robust in providing a transient acceleration.

\begin{figure}
\begin{minipage}[t]{0.5\linewidth}
\centering
\includegraphics[width=3.2in]{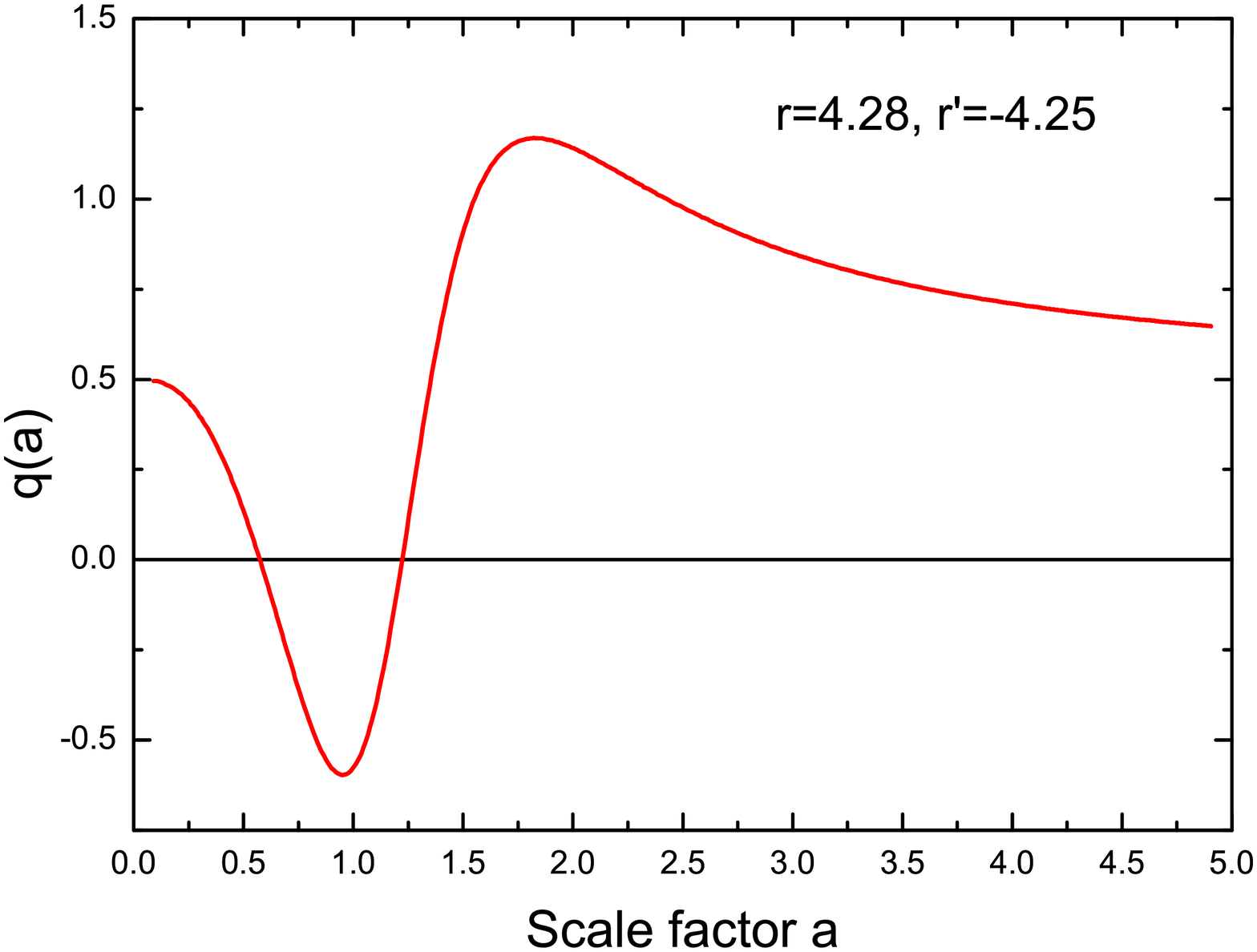}
\caption{\small The deceleration parameter $q$
        in Eq.(\ref{decelerationparameter2}) in the second model.}
\label{q2}
\end{minipage}
\hspace{0.5cm}
\begin{minipage}[t]{0.5\linewidth}
\centering
\includegraphics[width=3.2in]{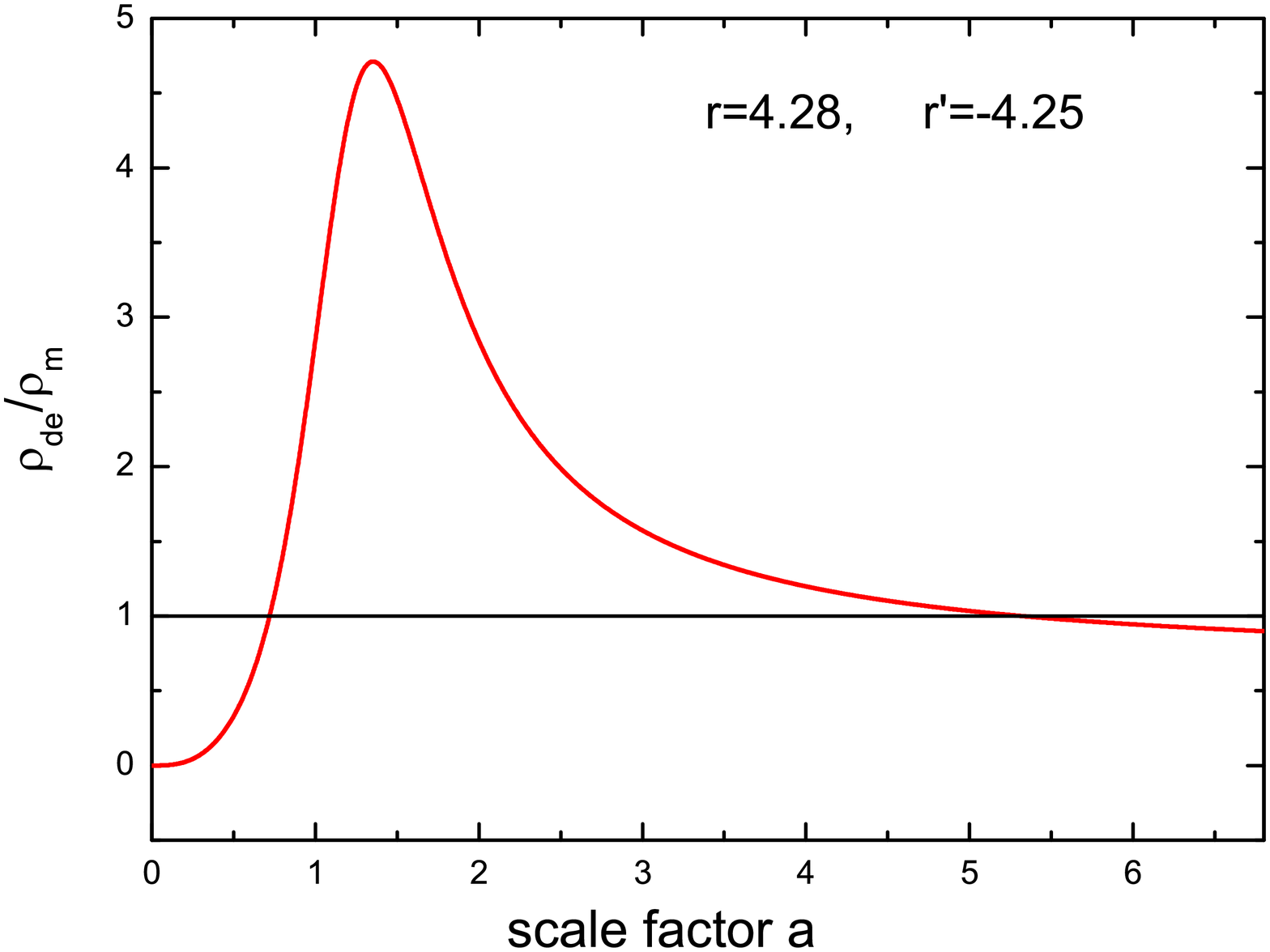}
\caption{\small The ratio $\rho_{de}/\rho_{m}$
              of Eq.(\ref{density 2}) in the second model.
              $\rho_{m}$ will dominate over $\rho_{de}$ in future.}
\label{r2}
\end{minipage}
\end{figure}

The above model can be extended
so that the universe in future is dominated by the matter $\rho_m$,
while retaining the return to deceleration.
We consider a dark energy density
\be \label{density 2}
\rho_{de}=\Omega_{de}\rho_c\frac{1+r+r'}{1+ra^3+r'a^{\epsilon (a)}},
\ee
where $r$ and $r'$ are constant parameters,
$\epsilon (a)$ is some function
of the scale factor $a$.
By the energy conservation, one obtains
\be \label{EOS1}
p_{de}=-\frac{\theta}{\Omega_{de}\rho_c(1+r+r')}\rho_{de}^2,
\ee
where
\be \label{coefficient1}
\theta \equiv 1+r'a^{\epsilon(a)}-\frac{1}{3}r'a\frac{d}{da}a^{\epsilon(a)}.
\ee
Then the deceleration parameter is
\be \label{decelerationparameter2}
q=\frac{1}{2}-\frac{3}{2}\frac{\Omega_{de}(1+r+r')\theta a^3}
{(1+ra^3+r'a^{\epsilon (a)})
[\Omega_m+(r+(1+r')\Omega_{de})a^3+\Omega_mr'a^{\epsilon (a)}]}.
\ee
This model admits a return to deceleration
as long as
 $\epsilon (a)>3$ for small $a$,
and  $\epsilon (a)< 3$ for large $a$.
For instance, taking $\epsilon (a)=(2a+4)/(a+1)$,
one has $\epsilon (a)\rightarrow 2$ for $a\rightarrow \infty$,
    and $\epsilon (a)\rightarrow 4$ for $a\rightarrow0$,
and the ratio
$\frac{\rho_{de}}{\rho_m}\rightarrow
\frac{\Omega_{de}}{\Omega_m}\frac{1+r+r'}{r}$ as $a \rightarrow \infty$.
A future matter domination is achieved if
 $\frac{1+r+r'}{r}< \frac{\Omega_m}{\Omega_{de}}$.
To be specific, we take $r=5.02$, $r'=-5.00$
 for  $\Omega_{de}=0.74, \Omega_m=0.26$,
yielding $\frac{\rho_{de}}{\rho_m}\rightarrow 0.704$ in the future.
Fig. \ref{q2} and Fig. \ref{r2}
demonstrate $q$ and $\frac{\rho_{de}}{\rho_m}$, respectively.
It is also checked that larger values of $r$ or $r'$
yield an earlier return to the deceleration expansion.
To be consistent with the error band $\pm 4\%$ for $ \Omega_{\Lambda}$
 by WMAP7 \cite{WMAP Collaboration},
the allowed value of the parameters
should be constrained to $r+r'< 0.04$.
In confronting  the joint data of SNIa, BAO and CMB,
the $\chi^2$ of this mode with $r=4.28$ and $r'=-4.25$ is also obtained
and shown (in dots) in Fig.\ref{chi}.
Its minimum is $\chi^2=543.60$ at $\Omega_m=0.276$.
Its corresponding likelihood ${\cal{L}}$ is $0.69$ times that of $\Lambda$CDM.
Thus, this second model is also close to $\Lambda$CDM   by statistical analysis,
although its is not as good as the first model.

\begin{figure}[htbp]
\centering
\includegraphics[width=0.60\columnwidth]{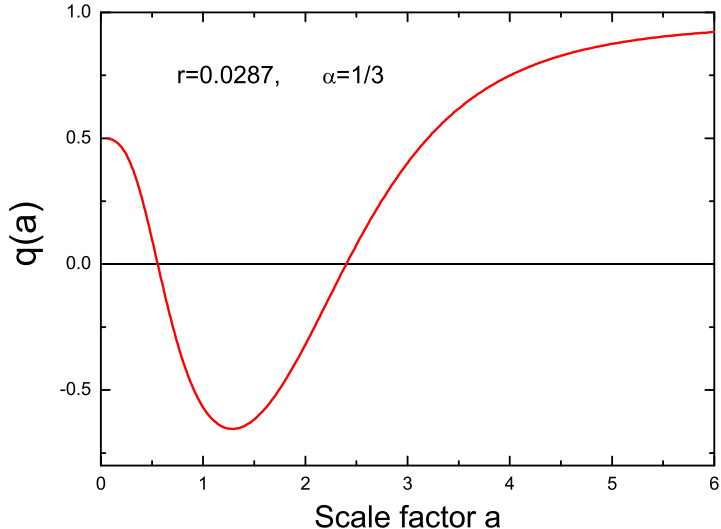}
 \caption{  \small The deceleration parameter $q$
 in Eq.(\ref{deceleration parameter3}) in the third model.
\label{q3}
 }
\end{figure}

In the above two models, the dark energy have a negative pressure $p_{de}<0$.
In fact,
the first model can be generalized
so that a positive pressure $p_{de}>0$ for $a\gg 1$ can be achieved.
Consider
\be \label{de density5}
\rho_{de}=\Omega_{de}\rho_c\frac{1+r}{1+ra^{3(\alpha+1)}},
\ee
where $\alpha$ is a positive constant.
In the limit $\alpha\rightarrow 0$,
this reduces to Eq.(\ref{de density}) of the first model.
From the energy conservation, the pressure is given by
\be \label{pressure 5}
p_{de}=\alpha \rho_{de}-\frac{1+\alpha}{\Omega_{de}\rho_c(1+r)} \rho_{de}^2,
\ee
consisting  of two terms.
When $a\gg 1$,
one has
$\rho_{de}\ll  \Omega_{de}\rho_c (1+r)  \frac{\alpha}{1+\alpha}$,
and $p_{de}\simeq \alpha \rho_{de}$,
which is positive.
From Eqs.(\ref{de density5}) and (\ref{pressure 5})
follows the deceleration parameter
\be \label{deceleration parameter3}
q=\frac{1}{2}  -\frac{3}{2}\frac{\Omega_{de}(1+r)(1-\alpha ra^{3\alpha+3})a^3}
{(1+ra^{3\alpha+3})[\Omega_m(1+ra^{3\alpha+3})+\Omega_{de}(1+r)a^3]},
\ee
which is shown  in Fig. \ref{q3}  for $\alpha=1/3$ and  $r=0.0287$.
When $a\gg 1$,   the matter will dominate, similar to the second model.
This dark energy can also be realized by a scalar field $\phi$.
By calculation,
we obtain
\be \label{scalar2}
2\sqrt{ra^{3(1+\alpha)}}
=e^{\sqrt{6\pi G(1+\alpha)}\phi}-e^{-\sqrt{6\pi G(1+\alpha)}\phi},
\ee
and the potential
\be
\label{potential3}
V(\phi)=2\Omega_{de}\rho_c(1+r)
\left[\frac{1-\alpha}{(e^{\sqrt{6\pi G(1+\alpha)}
\phi}+e^{-\sqrt{6\pi G(1+\alpha)}\phi})^2}
+4\frac{1+\alpha}{(e^{\sqrt{6\pi G(1+\alpha)}\phi}+e^{-\sqrt{6\pi G(1+\alpha)}\phi})^4}\right].
\ee
In the special case $\alpha=0$,
 Eq.(\ref{potential3}) reduces to Eq.(\ref{potential2}) of the first model.

In fact,
Eq.(\ref{pressure 5}) can be further extended
into the following most general form:
\be \label{eos2}
p_{de}=\alpha\rho_{de}-\beta\rho_{de}^m,
\ee
where $\alpha, \beta>0$ and $m>1$ are constant.
By the energy conservation,
Eq.(\ref{eos2}) yields
\be \label{density3}
\rho_{de}=\Omega_{de}\rho_c(\frac{1+r}{1+ra^{3(\alpha+1)(m-1)}})^{\frac{1}{m-1}},
\ee
where the parameter $r\equiv \frac{\sigma(\alpha+1)}{\beta}$,
and $\sigma $ is an integral constant and can be fixed by the initial condition
$\rho_{de}|_{a=1}=\Omega_{de}\rho_c$.
The deceleration parameter is
\be  \label{q4}
q(a)=\frac{1}{2}-\frac{3}{2}
\frac{\Omega_{de}
(1-\alpha ra^{3(\alpha+1)(m-1)})a^3 }{{  (1+ra^{3(\alpha+1)(m-1)})
[\Omega_{de}a^3+\Omega_m(\frac{1+ra^{3(\alpha+1)(m-1)}}{1+r})^{\frac{1}{m-1}}]}}.
\ee
When $m=2$, this general model   reduces to  the third model,
and reduces to the first model  if further $\alpha= 0$.
In the limit $\rho_{de}\ll(\alpha/\beta)^{1/m-1}$,
Eq.(\ref{eos2}) reduces to $p_{de} \simeq \alpha\rho_{de}$, a barotropic gas.
The general model in Eq.(\ref{density3})
has the following asymptotic behavior:
in the limit  $a\rightarrow0$,
$\rho_{de}\rightarrow (\frac{\alpha+1}{\beta})^{1/(m-1)}$
like the cosmological constant,
and,   in the limit $a\gg1$,  $\rho_{de}\rightarrow a^{-3(1+\alpha)}$,
which mimics a barotropic gas.
As we have checked by detailed computations,
shown in Fig. \ref{q4} for  $\alpha=1/3$ and $m=2.5$,
the general model  also admits a transient accelerating expansion,
in which  the matter will dominate for $a\gg 1$,
similar to the second and the third models.
Ref.\cite{Stefancic} also discussed
a special case of $\alpha=-1$ of Eq.(\ref{eos2})
in a different context.
\begin{figure}[htbp]
\centering
\includegraphics[width=0.60\columnwidth]{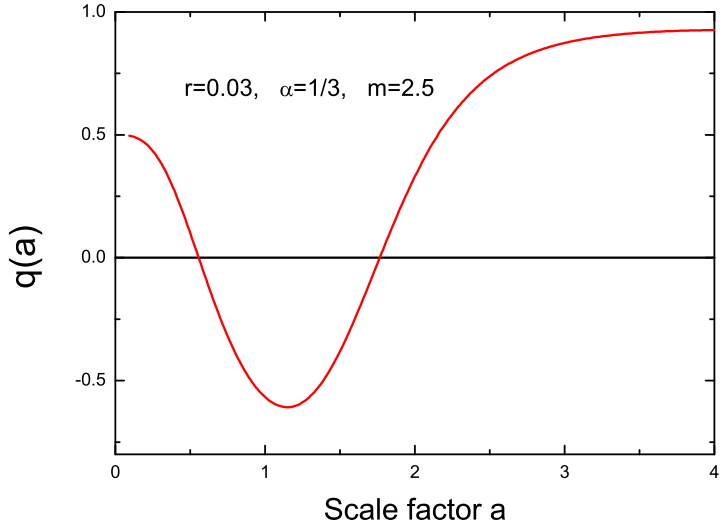}
 \caption{\label{model4}  \small The deceleration parameter $q$
 in Eq.(\ref{density3}) in the general model.
\label{q4}
 }
\end{figure}

\begin{center}
\textbf{3 Conclusion and Discussion}
\end{center}

We have demonstrated, by explicit model constructions, that
the current cosmic acceleration
driven by some dynamical dark energy can be transient,
and can transit smoothly into re-deceleration.
Four specific models have been examined in details,
each being a special case of the most general one
with the equation of state:  $p_{de}=\alpha\rho_{de}-\beta\rho_{de}^m$.
In all the models,
the dark energy behaves like a barotropic gas with
$\rho_{de}\rightarrow a^{-3(\alpha+1)}$ with $\alpha\ge 0$ for $a\gg 1$,
and the total cosmic energy can be
dominated either by $\rho_{m}$ as in the last three models,
or by $\rho_{de}$ as in the first model.
Our dark energy models can be realized by a scalar field $\phi$.
In two cases, by analytical integration,
we have also obtained the explicit expressions of scalar field potential $V(\phi)$,
which is as simple as a combination of the exponential functions of $\phi$,
and approaches to $V(\phi) \propto e^{-2\sqrt{6\pi G}\phi}$ for $a\gg 1$.
This function $V(\phi)$ differs from
the previous exponential type of potential in literature.

The interesting feature of these models is that
the dark energy density will always $\rho_{de}\rightarrow $ const,
subdominant to the matter during earlier stages.
This is an improvement over the previous models
of  transient acceleration that employed
the exponential type of scalar field potentials.
Moreover, in all our models
the fraction of dark energy at $a\ll 1$ is very close to
the value of the cosmological constant $\Omega_{\Lambda }$ in $\Lambda$CDM,
and is within the error band from WMAP observations.
Besides, the joint likelihood analysis
also shows that the transient acceleration models
can be as robust as $\Lambda$CDM
in confronting the observational data of SN Ia, CMB, and BAO.
Therefore, our models can be further incorporated
into the framework of the standard Big Bang cosmology
to achieve the possible transient acceleration.

\

\textbf{Acknowledgements}
Y. Zhang's research work has been supported by
the CNSF No.11073018, 11275187, SRFDP, and CAS.

\end{document}